# Microstructure and wear resistance of Fe-Cr-C-Mo-V-Ti-N hardfacing layers

Won Chol Son[1], Yong Gwang Jong[1,*], Myong Chol Pak[2,**], Jin Song Ma[3]

1 Faculty of Materials Engineering, Kimchaek University of Technology, Pyongyang, Democratic People's Republic of Korea

2 Department of Physics, Kim Il Sung University, Ryongnam Dong, Taesong District, Pyongyang, Democratic People's Republic of Korea

3 Faculty of Electrical Engineering, Kimchaek University of Technology, Pyongyang, Democratic People's Republic of Korea

* Corresponding author. E-mail address: jongyonggwang@163.com

** E-mail address: myongcholpak@163.com

## Abstract

In this paper, to improve wear resistance of components such as screws under severe friction-wear, Fe-Cr-C-Mo-V-Ti-N hardfacing coatings were further developed. The hardfacing coatings were acquired by shielded manual arc welding (SMAW) method. The ferroalloys added into the coating flux of the hardfaced electrode were jointly nitrided. The microstructure of the coatings was carried out using X-ray diffraction(XRD), optical microscope(OM), field emission scanning electron microscope (FESEM) and energy dispersive Xray spectrometry (EDS). In addition, FactSage 7.0 software was employed to calculate the equilibrium phase diagram of the hardfacings. The wear resistance was performed on a pin-on-disc machine. The Fe-Cr-C- Mo-V-Ti-N hardfacings exhibited higher wear resistance than cladding layer without nitrides.

## 1 Introduction

Fe-Cr-C coatings with high wear resistance have attracted great attention in the field of surface engineering[1]. These coatings generally have a composition in which primary $M_7C_3$-type carbides are formed in the welding deposition process, and the good wear resistance is due to the hard carbides[2]. But the coarse primary $M_7C_3$ often causes cracks during the solidification process, which may reduce the abrasive resistance of high chromium Fe-Cr-C hardfacing layers[3].

So the researchers studied the work to refine the primary carbide[4]. The different carbide-forming elements such as titanium, vanadium, niobium and tungsten were added to refine the primary carbide and to improve the wear resistance of high chromium Fe-Cr-C alloy[5–17].

However, the studies mentioned above have been of limited success[18]. Nowadays, researchers are studying the microstructure and properties of Fe-Cr-C alloys by adding nitrogen[19]. Niobium, vanadium and titanium are elements that have strong combination with C or N. It is found that these alloying elements formed the complex carbonitrides to refine the primary carbides and to improve the wear resistance of the coatings[20-23].

But the effect of jointly addition of nitrogen and nitride-forming elements such as vanadium, titanium and molybdenum in high chromium Fe-Cr-C hardfacing coatings is rarely found.

In this paper, to improve abrasive wear resistance of Fe-Cr-C hardfacing alloy, ferro-alloys such as ferrochromium, ferrovanadium, ferromolybdenum and ferrotitanium were jointly nitrided and tentatively added into coating flux of the hardfacing electrode. The present investigation is aimed at the understanding of carbonitride precipitates and wear resistance in the hardfacing alloy which is developed by SMAW method. Here, this new method makes the microstructure refined and wear resistance enhanced. Such studies are likely to provide better support to the reproduction of components such as screws under severe friction-wear condition.

## 2 Experimental materials and methods

In this study, the experimental diagram is shown in Fig. 1.

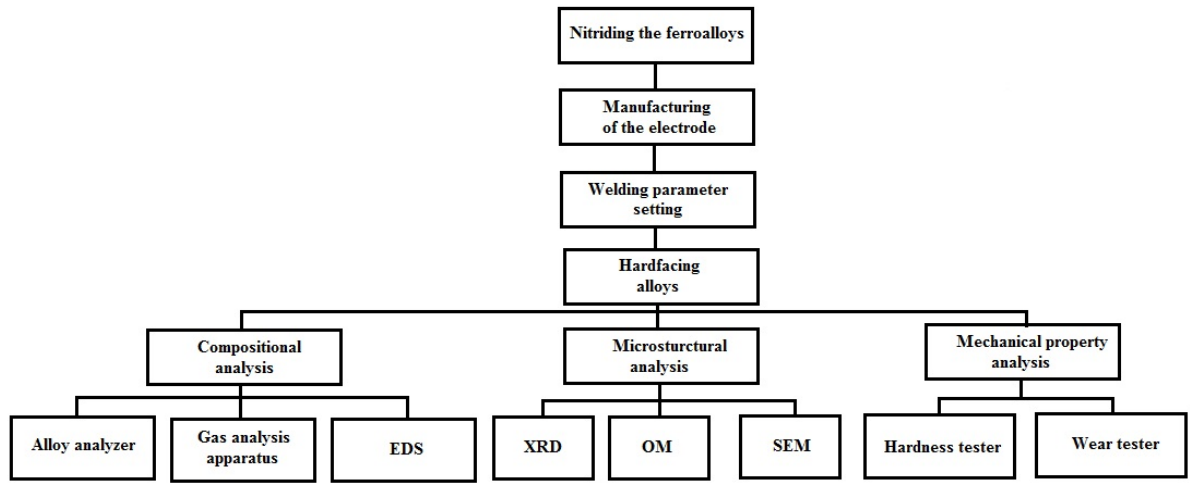

Fig.1 Experimental diagram

## 2.1 Experimantal materials

The core of the build-up electrode is Q235 with a diameter of 5mm (Table 1).

**Table 1 Chemical composition of Q235 steel (wt.%)**

| Element | C | Si | Mn | P | S |
|---|---|---|---|---|---|
| Content | ≤0.20 | ≤0.35 | ≤1.40 | ≤0.03 | ≤0.03 |

The compositions of the ferroalloys and the coating flux are shown in Table 2 and 3, respectively.

**Table 2 Chemical composition of the ferroalloys (wt.%)**

| Ferro-alloy | Cr | Ti | V | Mo | C | Si | P | S |
|---|---|---|---|---|---|---|---|---|
| Fe-Cr | 60 | - | - | - | 6.5 | 3.0 | 0.07 | 0.04 |
| Fe-Ti | - | 23 | - | - | 0.2 | 4.5 | 0.05 | 0.06 |
| Fe-V | - | - | 60 | - | 0.7 | 3.0 | 0.20 | 0.10 |
| Fe-Mo | - | - | - | 60 | 0.5 | 0.8 | 0.15 | 1.0 |

**Table 3 Composition of the coating flux (wt.%)**

| marble | fluorite | feldspar | graphite | Fe-Cr | Fe-V | Fe-Mo | Fe-Ti | Fe-Mn |
|---|---|---|---|---|---|---|---|---|
| 11 | 5 | 3 | 6 | 40 | 15 | 10 | 8 | 2 |

Nitriding the ferroalloys was performed as followed. The ferro-alloys mixed at a constant ratio (Table 2) were stretched with a thickness of about 5 to 7 mm on the surface of the container and then nitrided in a nitriding furnace. Firstly, after power was supplied to the nitriding furnace, it was heated to 450 °C and then nitrogen gas was injected into it. Subsequently, the furnace was maintained at 600 °C for 10 h, turned off the power, and provided nitrogen gas while being slowly cooled down to 300°C. The slag system is $CaO$-$CaF_2$-$SiO_2$. The coating flux includes 25% slag formers and 75% ferroalloys. The outer diameter of the electrode is 7.5mm.

Fig. 2 shows the schematic diagram of SMAW method. The coatings were deposited onto the steel substrates (Q235) with dimensions of 200 mm × 100 mm × 10 mm.

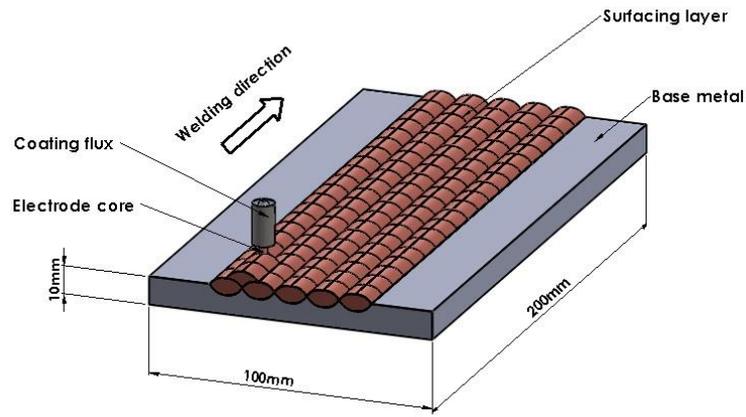

Fig.2 Schematic diagram of SMAW

## 2.2 Experimental methods

Before hardfacing, the steel substrates were ground and then cleaned with acetone. The hardfacing electrode was dried at 250°C for 2 h in a drying furnace. Fig. 3 shows the manufacturing process of the samples.

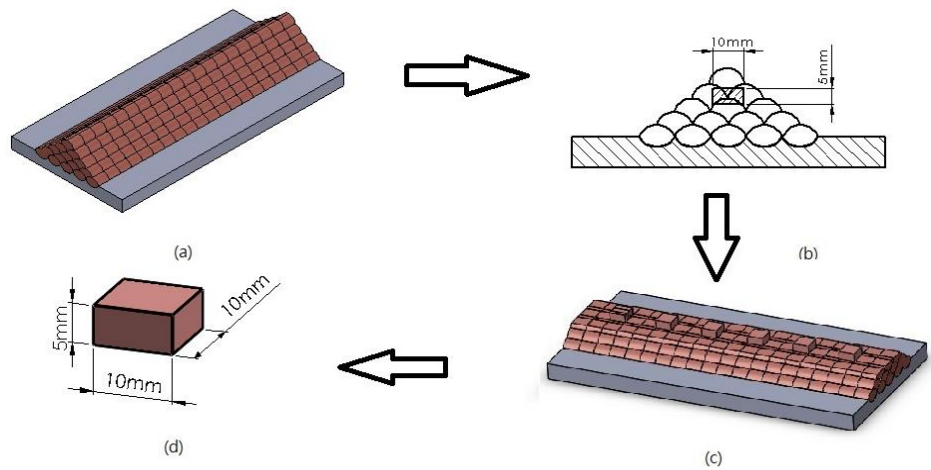

Fig. 3 Manufacturing process of the samples;
(a) hardfacing layer, (b) arrangement of the sample,
(c) manufacturing of the sample, (d) size of the sample

Using SMAW method, five layers of hardfacing were deposited onto the surface of each steel substrate to get the homogeneous specimen. The welding parameters and the total thickness of the layer are listed in Table 4.

Table 4 Hardfacing processing parameters and the total thickness of the layer

| Current(A) | Voltage(V) | Speed(cm/min) | Layer Thickness(mm) |
|---|---|---|---|
| 180~200 | 20~25 | 8~10 | 15 |

Specimens with dimensions of 10 mm ×5 mm ×10 mm were cut from the longitudinal section of the weld layer. The specimens were ground with silicon carbide abrasive paper from 80 up to 1200 grit and were polished with diamond compound polishing paste. The chemical composition of the coatings was observed by a QSN750 alloy analyzer and nitrogen analysis was carried out by EGMA-1300 gas analysis apparatus. The phase structure of the coatings was determined by X-ray diffraction (XRD) using a D/max-2500/PC instrument. After etching with a solution of 5g of $FeCl_3$ +10ml of $HNO_3$+3ml of HCl+87ml of ethyl alcohol, the microstructure of the samples was observed by an OLYMPUS–DSX500 optical microscope and a ZEISS field emission scanning electron microscope(FESEM). The composition of each phase was determined by energy dispersive X-ray spectrometry (EDS).

Wear test was performed on an MMU-10G pin-on-disc machine at room temperature (20°C). 40Cr steel was selected as the abrasive material. The cylindrical pin specimen was the dimension of ∅ 4.8mm×13 mm. The specimen was tested under a normal load of 300N and the rotating speed of it was 220rpm. An electronic balance was used to weight the mass loss of the specimens per 10 min. The wear resistance of the hardfacing layers deposited with nitrided ferroalloys was compared to one with non-nitrided ferroalloys. To reduce experimental error, three samples of each group were prepared for the test. The measured value is average of the measured values from three samples.

FactSage7.0 (FSstel-FactSage steel database) was used to calculate the phase diagram of hardfacing coatings.

### 3 Results and discussion
#### 3.1 Microstructure and phase structure

The hardfacing coatings with non-nitrided ferroalloys were marked as S1 and other hardfacing coatings with nitrided ferroalloys were marked as S2. Table 5 shows the chemical compositions of the hardfacing coatings without and with nitrided ferroalloys. It was measured at the top surface of the hardfacing layers. The result shows the high transition ratio of molybdenum. This resulted in a lower affinity of Mo with oxygen compared to the V, Ti elements. Thus, during welding process, most of Mo could be transferred to the hardfacing layer.

Table 5 Chemical compositions of the hardfacing alloys (wt.%)

| specimens | C | Cr | Mo | Mn | Ti | V | N |
|---|---|---|---|---|---|---|---|
| S1 | 1.86 | 12.1 | 3.1 | 0.57 | 0.09 | 2.25 | - |
| S2 | 1.84 | 12.5 | 3.2 | 0.58 | 0.10 | 2.85 | 0.28 |

Fig. 4 shows the optical micrograph (OM) and XRD results from the longitudinal section of S1. It was seen that the coating consisted of α-Fe, (Fe, Cr)$_7$C$_3$, austenite, martensite and V$_8$C$_7$. Fig. 4a represents the dispersed distribution of fine grains with irregular polygonal morphology in the matrix and grain boundary. These grains are small in size and mostly in the 1-2 μm range. Fig. 5 presents the microstructure and XRD patterns of S2. The X-ray phase examination of this hardfacing reveals that S2 mainly consists of α-Fe, (Fe, Cr)$_7$C$_3$, austenite, martensite, V$_8$C$_7$ and MX(M=V, Ti; X=C, N)-type carbonitrides (Fig. 5b). Meanwhile, it is found that S2 has a finer grain than S1. So the complex carbonitrides have an obvious effect on grain refining[23].

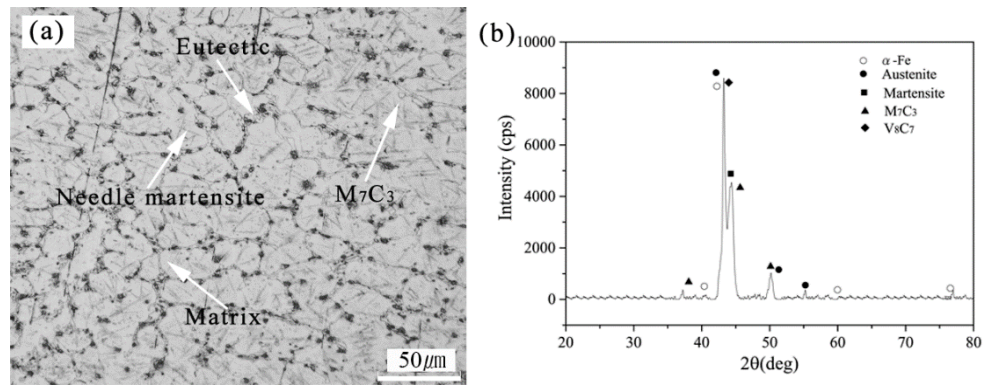

Fig.4 Microstructure(a) and XRD results(b) of the S1

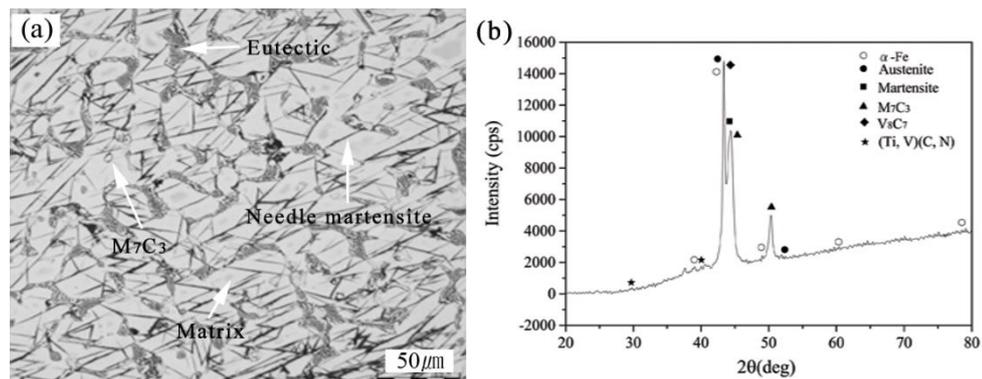

Fig.5 Microstructure(a) and XRD results(b) of the S2

Fig. 6 shows the SEM and EDS results of the longitudinal section of S2. The EDS analysis of Fig. 6a position 1 shows the presence of vanadium, Fe, chromium, molybdenum, titanium, carbon and nitrogen peaks. So it is determined that these precipitate particles homogeneously distributed in S2 are complex carbonitrides of V and Ti[23]. The EDS analysis results of position 2 and 3 reveal the presence of chromium, Fe, carbon, vanadium and molybdenum peaks (Figs. 6c and 6d). Fig. 5 and EDS analysis results verify that the area 2 is (Fe, Cr)$_7$C$_3$ carbide and the area 3 is eutectic carbides.

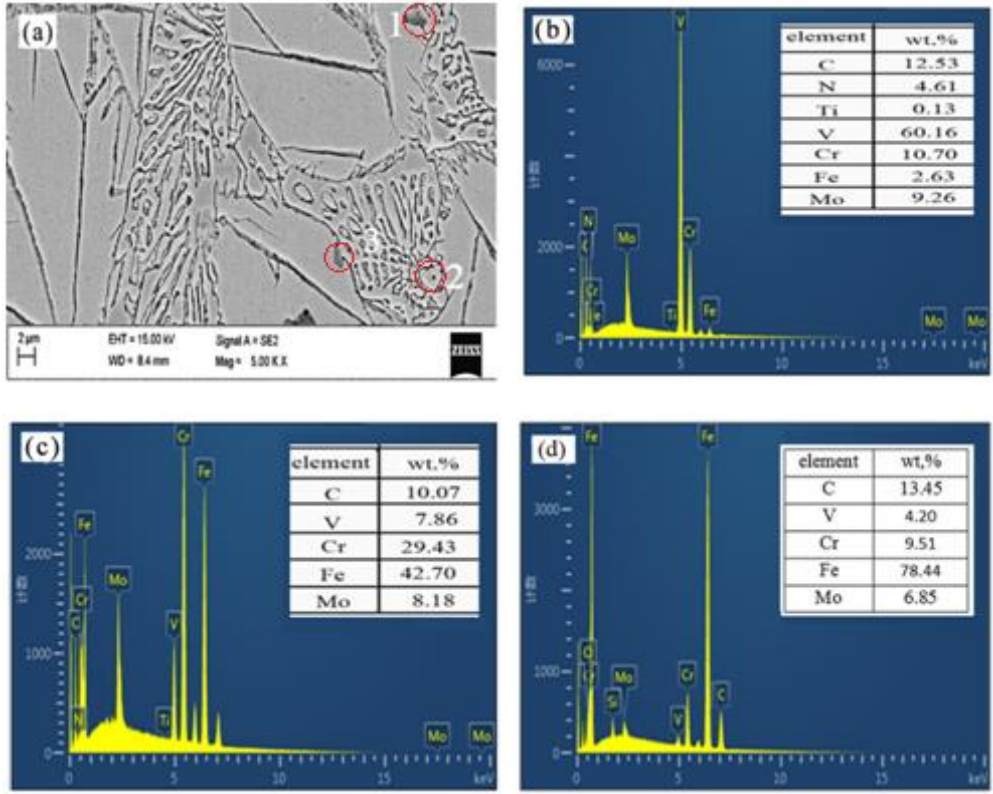

Fig.6 SEM image(a) and EDS results(b, c, d) of hardfacing layer;
(b) position 1 in SEM, (c) position 2 in SEM, (d) position 3 in SEM

Table 6 gives the hardness of the two samples. Compared with S1, S2 has a 3 HRC increase in the hardness value. It is referred to the fact that comparative with S1, the complex carbonitrides ((Ti, V)(C, N)) in S2 were precipitated from the hardfacing alloy (Fig. 5b). It was found that the complex carbonitrides can precipitate out with very fine size and make a great secondary hardening effect on the matrix[20].

**Table 6 Hardness of the hardfacing alloys**

| specimens | Hardness (HRC) |
| --- | --- |
| S1 | 56 |
| S2 | 59 |

Fig.7 shows the increasing rate in the weight of ferro-alloys which were jointly nitrided in the individual containers. It could be appeared that the weight increasing rate of Fe-V is larger than Fe-Ti. Because the amount of ferrotitanium is less than ferrovanadium in ferroalloys added into the coating flux (Table 3) and titanium (23%) is less contained in ferrotitanium. That is why the content of TiN will be low relatively in the hardfacing coatings.

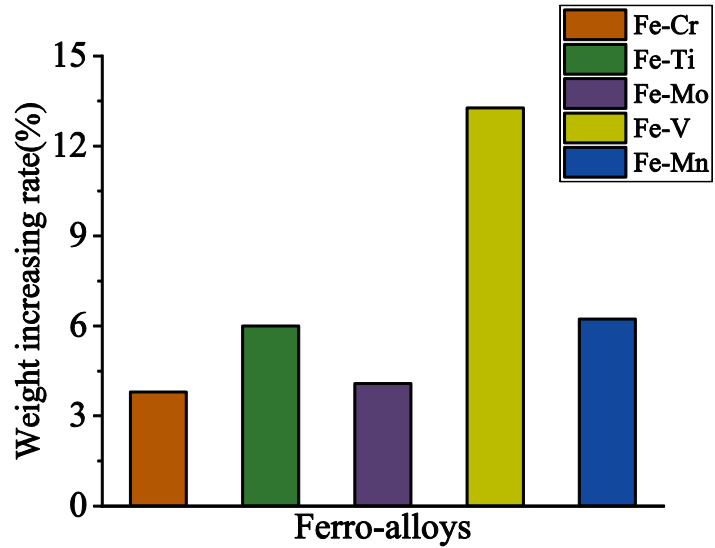

Fig. 7 Weight increasing rate of the individual ferroalloys after nitridation

So it will be insufficient that titanium reacts with carbon and nitrogen to form TiC carbide or $Ti_x(C, N)_y$ carbide in welding metallurgical reaction. On the other hand, most of ferrotitanium will be exhausted in welding process because titanium has the greater affinity with oxygen than vanadium and molybdenum. In addition, V is more likely to be dissolved in matrix due to its better wettability to Fe than Ti. It is consistent with the EDS results (Fig. 6b). Therefore, the possibility of forming abundant V-rich carbides or carbonitrides during the welding metallurgical reaction can be greatly increased.

**3.2 Wear resistance**

Fig.8 shows the relationship between mass loss and wear time of the coatings. It can be seen that mass loss of S2 was lower than that of S1 at the same condition. Investigated hardfacing layers were consisted of complex carbides and carbonitrides of V, Mo and Ti. Carbonitrides are homogeneously distributed on the boundary and in the matrix, which contributes to the strengthening of the hardfacing layer. Therefore, S2 with fine carbonitrides presented better rigidity than S1 with coarse carbide phases. In conclusion, S2 showed better abrasion resistance due to the combination effect of substrate and hardfacing coatings, resulting in less matrix removal. So, the wear damage of S2 was reduced, and the mass loss became slower.

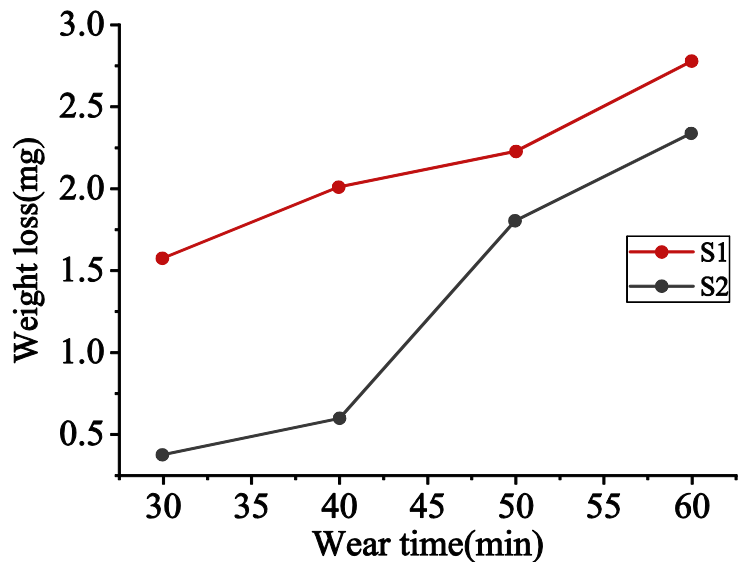

Fig.8 Relationship between mass loss and wear time of coatings

The SEM morphologies of the worn surface after wear test (after 60 min) are shown in Fig. 9. The worn surface of the S1 is more rough than the S2, with relatively numerous adhesive craters, deep ploughing grooves and a lot of detached wear debris as shown in Fig. 9.

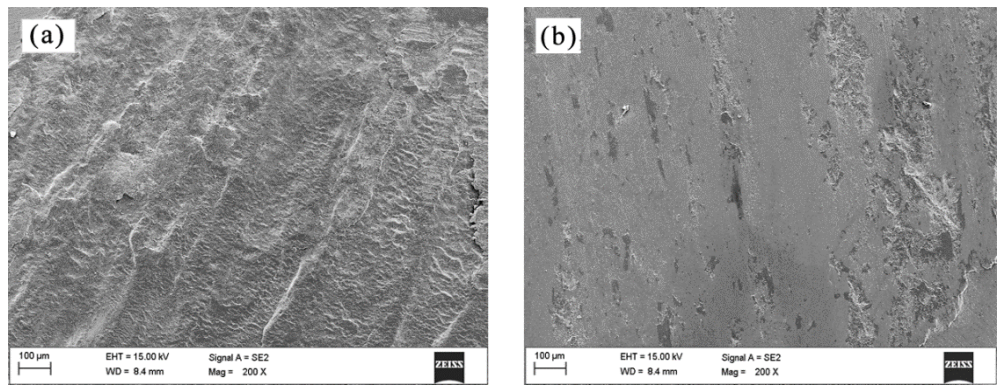

Fig.9 Worn surface morphologies of hardfacing metals:
(a) S1;　(b) S2

Compared with the hardfacing layer with nitrides, hardfacing layer without nitrides is lower bonding with matrix. In the friction-wear process of hardfacing layers without nitrides, the large carbides are easily exfoliated from the matrix, causing microcutting and microploughing scratches. The existence of well-distributed carbonitrides helps minimize the wear, hence, the layer of the S2 displays the best wear-resistance.

### 3.3 Thermodynamic calculation of carbonitrides precipitation

In order to analyze the precipitation rule of the phases in surfacing layer and the possibility of carbides and carbonitrides forming during the surfacing solidification process, the equilibrium phase diagram and the mass fraction of all solid phases of Fe-Cr-C-Mo-V-Ti-N hardfacing metal were calculated (Figs.10 and 11). As shown in Fig. 10, the eutectic temperature is located at about 3.8%C. Thus, the microstructure of hardfacing metal with 1.68%C is a hypoeutectic structure.

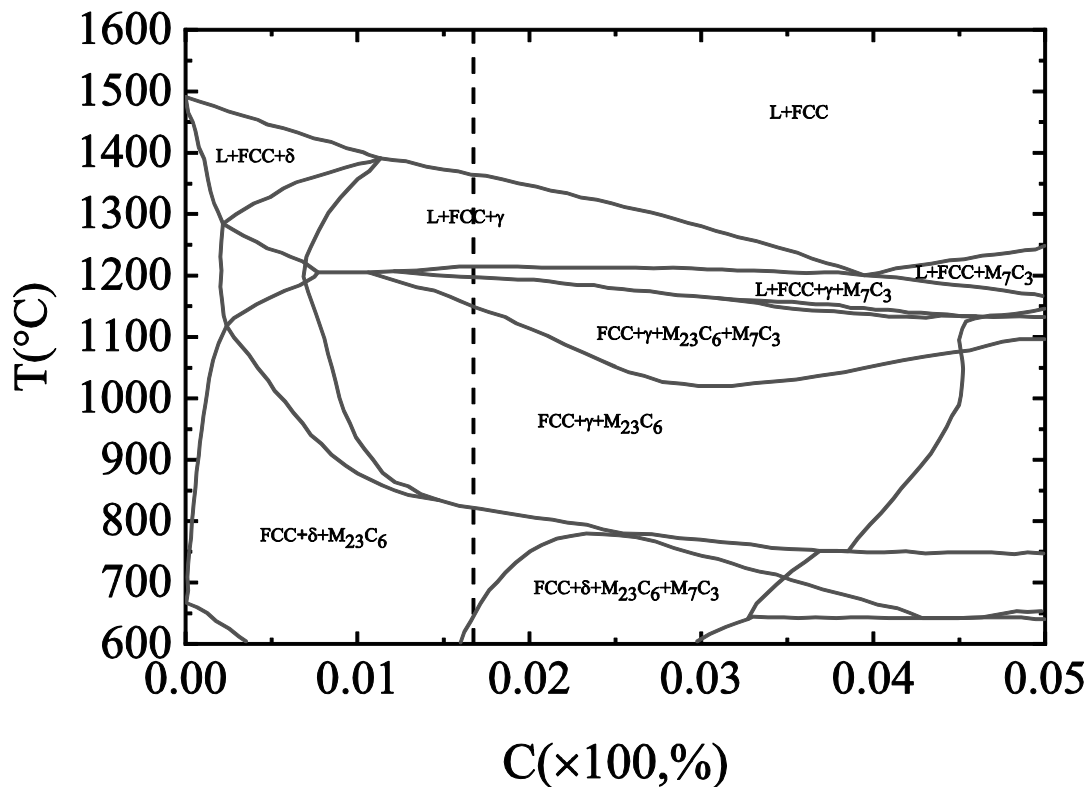

Fig. 10 Equilibrium phase diagram of the Fe- Cr-C-Mo-V-Ti-N hardfacing metal

Here, when the temperature is attained to 1360°C, the primary austenite is initiated to precipitate from the liquid phase (red curve, Fig. 11). Subsequently, the $M_7C_3$ and $M_{23}C_6$ carbide precipitations occur at 1212°C and 1199°C, respectively (purple curve and green curve, Fig.11). The liquid disappears when the temperature is lower than 1190 °C.

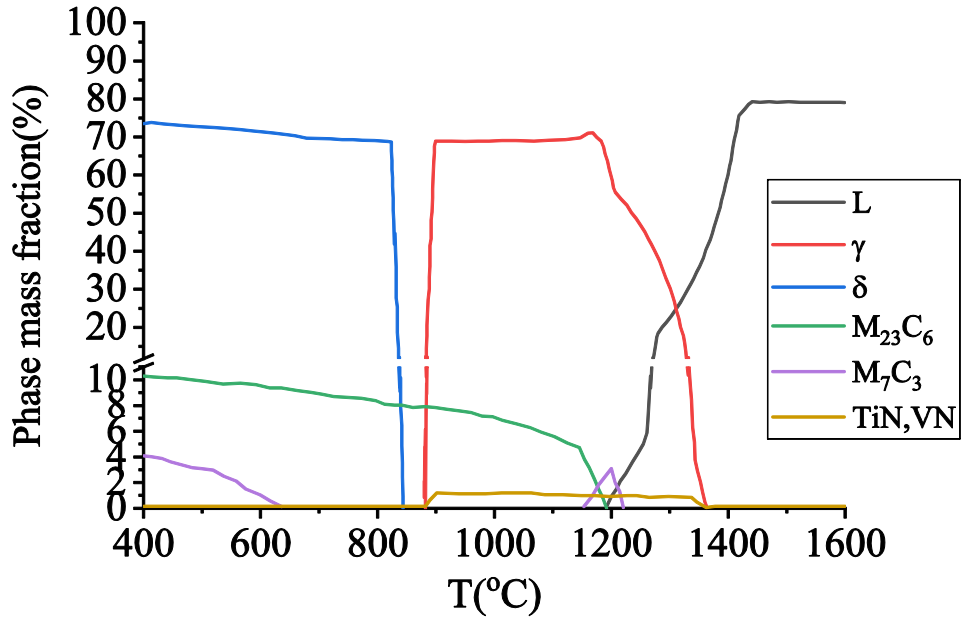

Fig. 11 Mass fraction of all solid phases of Fe-Cr-C-Mo-V-Ti-N hardfacing metal

And the mass fraction of austenite decreases rapidly at 850°C and the δ phase starts to form at 822°C (red curve and blue curve, Fig. 11). Meanwhile, from Figs. 10 and 11, it can be confirmed that precipitates of FCC lattice existed in the liquid zone will be nitrides of Ti and V (brown curve, Fig. 11). It shows that the nitrides of Ti and V have been already occurred in the liquid zone. In addition, it can be seen that TiN and VN may be existed from the liquid to the complete solidification (brown curve, Fig. 11). The microstructure of Fe-Cr-C-Mo-V-Ti-N alloy is $δ+M_7C_3I+M_7C_3II+M_{23}C_6+MX(M=V, Ti; X=C, N)$.

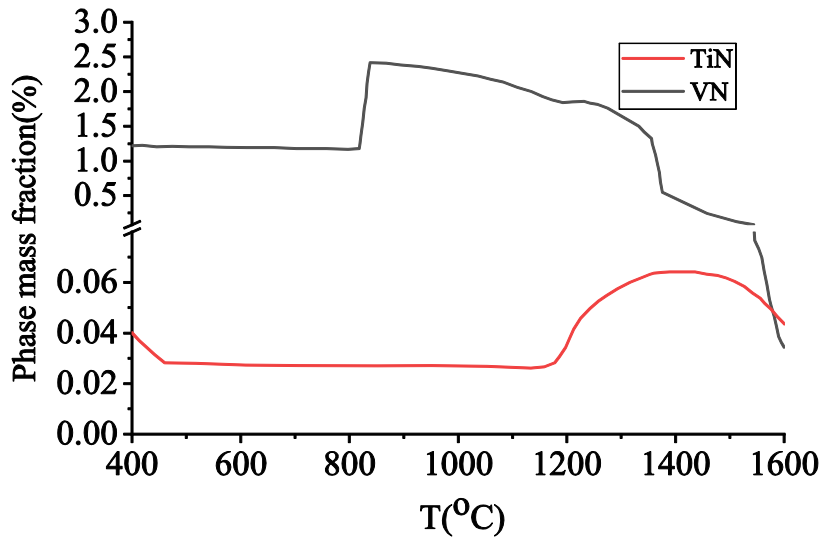

Fig.12 Mass fraction of TiN and VN according to the temperature

The solidification curves of TiN and VN in the Fe-Cr-C-Mo-V-Ti-N alloy with the temperature were calculated, which were shown in Fig. 12. As shown in Fig. 12, the amount of VN (1.25%) was larger than that of TiN (0.04%). This is obvious as shown in Fig. 6. Therefore, the precipitated (V,Ti)(C,N) carbonitrides are enriched with vanadium.

## 4 Conclusions

In this study, the microstructure and the wear resistance of hardfacing coatings (with and without nitrided ferroalloys) have been analysed. The complex carbonitrides ((Ti, V)(C, N)) were precipitated from the hardfacing alloy with nitrided ferroalloys. The carbonitrides were existed from the liquid to the complete solidification and enriched with vanadium. The complex carbonitrides can precipitate out with very fine size and make a great secondary hardening effect on the matrix. Thus, the hardfacing coatings with fine carbonitrides presented better abrasion resistance, resulting in less matrix removal. Therefore, the wear damage of that was reduced, and the mass loss became slower.

**Acknowledgements**

The authors would like to thank Hang Son Kim and Sang Yun Pak in the Faculty of Materials Engineering at Kimchaek University of Technology. This work is financially supported by Ministry of Science and Technical Development, Democratic People's Republic of Korea, Project No.5-11-32.